\documentclass[9pt,biorxiv]{lapreprint}

\newcommand{\paperTitle}{PyGPLA: A Python Package for Generalized Phase Locking Analysis}
\newcommand{\shortPaperTitle}{PyGPLA: A Python Package for Generalized Phase Locking Analysis}

\newcommand{\authorone}{Amir Khani}
\newcommand{\authortwo}{Christian Beste}
\newcommand{\authorsix}{Shervin Safavi}

\newcommand{\afillA}{Department of Computer Science, Amirkabir University of Technology, Tehran, Iran}
\newcommand{\afillC}{Cognitive Neurophysiology, Department of Child and Adolescent Psychiatry, Faculty of Medicine, TU Dresden, Dresden, Germany}

\newcommand{\afillF}{Department of Computational Neuroscience, Max Planck Institute for Biological Cybernetics, T\"ubingen, Germany}
\newcommand{\afillB}{Computational Neuroscience, Department of Child and Adolescent Psychiatry, Faculty of Medicine, TU Dresden, Dresden, Germany}
\newcommand{\afillH}{Donders Institute for Brain, Cognition, and Behaviour, Radboud University, Nijmegen, Netherlands}
\newcommand{\afillI}{German Center for Child and Adolescent Health (DZKJ), partner site Leipzig/Dresden, Dresden, Germany}

\newcommand{\cauthorEmail}{amir.khani@donders.ru.nl, research@shervinsafavi.org}
\newcommand{\cauthorInitials}{AK, SS}

\newcommand{\leadAuthor}{Khani}

\usepackage[version=4]{mhchem} 
\usepackage{siunitx}    
\usepackage{pdflscape}  
\usepackage{rotating}   
\usepackage{textgreek}  
\usepackage{gensymb}    
\usepackage[misc]{ifsym} 
\usepackage{orcidlink}  
\usepackage{listings}   
\usepackage{colortbl}   
\usepackage{tabularx}   
\usepackage{longtable}  
\usepackage{subcaption}
\usepackage{multirow}
\usepackage{snotez}     
\usepackage{csquotes}   

\DeclareSIUnit\Molar{M}

\title{\paperTitle}

\author[1, 2 \Letter]{\authorone}
\author[3, 4]{\authortwo}

\author[ \orcidlink{0000-0002-2868-530X} 5, 6 \Letter]{\authorsix}

\affil[1]{\afillH}
\affil[2]{\afillA}
\affil[3]{\afillC}
\affil[4]{\afillI}
\affil[5]{\afillB}
\affil[6]{\afillF}
\correspondence{\cauthorEmail}{\cauthorInitials}

\leadauthor{\leadAuthor}
\shorttitle{\shortPaperTitle}

\usepackage{hyperref}
\usepackage{cleveref}

\usepackage{newfloat}

\newcounter{stable}
\newcounter{tempTableCounter}

\crefname{table}{Table}{Tables}
\Crefname{table}{Table}{Tables}
\crefname{stable}{Supp. Table}{Supp. Tables}
\Crefname{stable}{Supp. Table}{Supp. Tables}
\crefname{slongtable}{Supp. Table}{Supp. Tables}
\Crefname{slongtable}{Supp. Table}{Supp. Tables}
\crefname{longtable}{Table}{Tables}
\Crefname{longtable}{Table}{Tables}

\DeclareFloatingEnvironment[fileext=los,name=Supplementary Figure]{sfigure}

\newcommand{\stab}[1]{%
  \hyperref[#1]{Supp. Table~\ref*{#1}}%
}

\AtBeginDocument{%
}

\crefname{equation}{Eq.}{Eq.}
\crefname{figure}{Fig.}{Fig.}
\Crefname{equation}{Eq.}{Eqs.}
\Crefname{figure}{Fig.}{Figs.}

\providecommand{\Crefname}[3]{}%
\crefname{sfigure}{Supp. Fig.}{Supp. Figs.}
\Crefname{sfigure}{Supp. Fig.}{Supp. Figs.}
\crefname{figure}{Figure}{Figure}

\newcommand{\mfig}[1]{\cref{#1}}
\usepackage{xfrac}
\usepackage{placeins}

\usepackage{xspace}

\usepackage[textsize = tiny,backgroundcolor=blue!8!white]{todonotes}
\newcommand{\cut}[1]{}

\newcommand{\shsc}[1]{}
\newcommand{\shsd}[1]{}
\newcommand{\shstd}[1]{}

\newcommand{\oltextd}[1]{} 

    \usepackage[			
    	backend=biber,      
            style=numeric-comp,   
    	natbib=true,		
    	hyperref=true,	    
    	alldates=year,      
            mincitenames=1,
            maxcitenames=1,
            sorting=none,
            backref=false
            ]{biblatex}
    
    \renewbibmacro{in:}{}
    
    \DeclareNameAlias{default}{family}
    \DeclareNameAlias{sortname}{default}
    \DeclareNameAlias{labelname}{default}
    \DeclareNameAlias{sortname}{family-given}
    
    \DeclareNameFormat{family}{%
      \usebibmacro{name:family}
      {\namepartfamily}
      {\namepartgiven}
      {\namepartprefix}
      {\namepartsuffix}%
      \usebibmacro{name:andothers}}
      
\usepackage{float}

\newcommand{\pygplaName}{\texttt{PyGPLA}}

\begin{document}
\maketitle


\begin{abstract}
Spike--field couplings measure the links between neuronal spiking and local field potential (LFP) and provide a tool to study how neural populations coordinate
activity across spatial and temporal scales. 
While modern high-density electrophysiology enables the simultaneous recording of numerous LFP channels and neuronal units, this massive scale poses significant statistical and computational challenges for conventional pairwise coupling analyses.
The number of channel--unit comparisons grows rapidly,
and a coherent population-level structure can be obscured by
high-dimensional coupling matrices. Generalized Phase Locking Analysis (GPLA)
addresses this challenge by organizing spike-triggered LFP phase information into a complex coupling matrix and extracting dominant low-rank modes.
These modes summarize whether a population is coupled to the field, which channels and units contribute to the coupling, and how their relative phases are organized.
Here, we present \pygplaName, an open-source Python package implementing GPLA for high-density electrophysiological recordings.
Designed for straightforward integration into existing neural data analysis pipelines, \pygplaName\ supports multivariate spike--LFP coupling analysis, from coupling estimation and interpretation to significance assessment using analytical and surrogate-based tests.
\end{abstract}

\section{Statement of need}

The increasing complexity of high-dimensional, multi-modal neural datasets necessitates the development of appropriately sophisticated analysis tools \citep{Stringer2024AnalysisMethods, Siegle2026LargeScaleElectrophysiology, einevoll2013modelling, stevenson2011HowAdvancesNeural, Newman2025ONIX}.
Simultaneous analysis of spike and LFP is among the very first pairs of high-dimensional and multi-modal signals in systems neuroscience, 
and their multi-modal analysis has been insightful about the multi-scale machinery supporting a wide range of cognitive functions \citep{einevoll2013modelling}. 
LFP oscillatory activity partly reflects
subthreshold processes shared by neural assemblies \citep{buzsaki2012origin, einevoll2013modelling, hagen2016hybrid,herreras2016local},
and the synchronization between such activities and spiking is hypothesized to coordinate neural populations and support diverse cognitive functions such as perception, attention and decision-making \citep{Fries2005,friesRhythmsCognitionCommunication2015, womelsdorfModulationNeuronalInteractions2007a, itoSaccaderelatedModulationsNeuronal2011, dwarakanath2023bistability}.
However, uncovering these mechanisms faces a major obstacle.
Commonly used spike–field coupling measures are pairwise \cite{zeitler2006assessing, ashida2010processing, vinck2010pairwise, vinck2012improved, jiang2015measuring, li2016unbiased, zarei2018introducing}, which makes them arguably suboptimal for modern multichannel recording systems \citep{buzsaki2004large, dickeySingleUnitStabilityUsing2009, junFullyIntegratedSilicon2017a, pesaranInvestigatingLargescaleBrain2018, juavinett2019chronically}.
As contemporary electrophysiological techniques allow simultaneous recordings from hundreds or even thousands of sites, pairwise analyses generate high-dimensional matrices whose size grows rapidly with the number of channels, making interpretable extraction difficult and limiting our ability to characterize large-scale collective dynamics linking spiking activity to mesoscopic LFP patterns \citep{ermentrout2001traveling, cruddasCorticalTravelingWaves2026, doi:10.1137/06066638X, zhangThetaAlphaOscillations2018, mullerCorticalTravellingWaves2018, dugueTravelingWavesScales2025}.

The Generalized Phase Locking Analysis \citep[GPLA, ][]{safavi2023uncovering} was developed to address
these challenges by providing an efficient multivariate framework along with statistical routines \citep{safaviUnivariateMultivariateCoupling2021} for
characterizing spike-field coupling at the population level.
However, the original algorithm was implemented in MATLAB,
limiting its accessibility. To bridge this gap, we introduce \pygplaName,
an open-source Python implementation of GPLA.
Aligning with the extensive use of Python in neuroscience \citep{GramfortEtAl2013a, muller2015python, peirce2009generating, ince2009python, krause2014expyriment, tennoe2018uncertainpy, freeman2015open, zeraati2022flexible, akam2022open, makowski2021neurokit2, goodman2008brian, behrad2025fast, Nouri2025, viejo2023pynapple}, \pygplaName\ provides a Python toolbox for studying population-level spike--field coupling in large-scale neural recordings.

\section{Functionality}
In this section, we describe the main functionality of \pygplaName. We first briefly present the mathematical formulation of GPLA, then detail the pre- and post-processing steps used to compute and interpret its coupling estimates, and finally explain how the package is organized to implement this workflow.

\subsection{Mathematical formulation of GPLA}

GPLA generalizes classical spike--field coupling (SFC) to multichannel recordings by casting
all spike--LFP pairwise couplings into a single complex-valued matrix and extracting its dominant
low-rank structure \citep{safavi2023uncovering}. For a given pair consisting of a single-unit and an LFP channel, let
$L_f(t)$ denote the narrowband analytic LFP at center frequency $f$, obtained by band-pass
filtering followed by the Hilbert transform, so that
$L_f(t)=a_f(t)e^{i\phi_f(t)}$, with instantaneous amplitude $a_f(t)$ and phase $\phi_f(t)$.
A straightforward definition of a complex SFC metric is
\begin{equation}
c(f)=\langle \lambda(t)L_f(t)\rangle
      = |c|e^{i\Phi_c}\,,
\end{equation}
where $\langle\cdot\rangle$ denotes averaging across time (and trials) and $\lambda(t)$ is the
instantaneous spike rate. The modulus $|c|$ reflects coupling strength and the argument
$\Phi_c$ is the dominant LFP phase of spiking. This statistic is closely related to the
phase-locking value (PLV) but differs by amplitude-weighting through $a_f(t)$ and by the
absence of spike-rate normalization \citep{safavi2023uncovering}.

In multivariate recordings, let $L_f^{n,(k)}(t)$ be the analytic LFP for channel $n\in\{1,\dots,N_c\}$
in trial $k\in\{1,\dots,K\}$, and let $\{t_j^{m,(k)}\}_{j=1}^{N_m^{(k)}}$ be the spike times of unit
$m\in\{1,\dots,N_u\}$. The (unnormalized) coupling-matrix estimator
$\widehat{\mathbf C}(f)\in\mathbb{C}^{N_c\times N_u}$ is
\begin{equation}
\widehat{\mathbf C}_{n,m}(f)=\sum_{k=1}^{K}\sum_{j=1}^{N_m^{(k)}} L_f^{n,(k)}\!\left(t_j^{m,(k)}\right)\,,
\end{equation}
i.e., each entry sums the analytic LFP evaluated at all spike times \citep{safavi2023uncovering}.
A normalized matrix $\mathbf C(f)$ is obtained by scaling columns by the total spike count
$N_m^{\mathrm{tot}}=\sum_k N_m^{(k)}$ \citep[e.g., $1/N_m^{\mathrm{tot}}$ for phase-only PLV-type scaling,
or $1/\sqrt{N_m^{\mathrm{tot}}}$ for unit-variance scaling used by the analytical significance test; see below and ][for more details]{safaviUnivariateMultivariateCoupling2021}.

GPLA summarizes the coupling matrix, $\mathbf C(f)$, via singular value decomposition (SVD)
\begin{equation}
\mathbf C = \mathbf U\mathbf D\mathbf V^{H}=\sum_{r=1}^{p} d_r\,\mathbf u_r \mathbf v_r^{H},
\qquad p=\min(N_c,N_u)\,,
\end{equation}
and focuses on the leading mode \(\mathbf{C} \approx d_1 \mathbf{u}_1 \mathbf{v}_1^H\).  
The largest singular value \(d_1\) is the generalized phase-locking value (gPLV), which quantifies the strength of population-level coupling, while \(\mathbf{u}_1\) (LFP vector) and \(\mathbf{v}_1\) (spike vector) encode the dominant patterns of the coupling in LFP and population spiking. The magnitudes of the vector entries indicate the relative contribution of each channel/unit to the coupling, while the phases encode the relative timing between spiking and LFP oscillations.  
Because the factorization is invariant to a global complex phase, GPLA resolves this ambiguity by fixing the global phase convention. Specifically, we choose the global phase of the LFP vector such that \(\arg\left(\frac{1}{N_c}\sum_{n=1}^{N_c} u_{1,n}\right) = 0\). Consequently, the mean phase of the spike vector, \(\arg\left(\frac{1}{N_u}\sum_{m=1}^{N_u} v_{1,m}\right)\), reflects the overall spike-LFP phase lag, summarizing the temporal relationship between spiking and field activity at frequency \(f\) \citep{safavi2023uncovering}.

\subsection{Pre- and post-processing}

The construction and interpretation of the coupling matrix involve three groups of operations. First, raw LFP signals must undergo frequency-domain filtering and conversion to an analytic-signal or phase representation before being supplied to \pygplaName. These experiment-specific steps are performed upstream, while optional whitening is implemented within \pygplaName. Second, an optional spike-count normalization is applied to the coupling entries, which controls the variance structure of the estimator \citep[which is essential for theoretical statistical analysis; see ][for more details]{safavi2023uncovering, safaviUnivariateMultivariateCoupling2021}. Third, after the SVD, a set of post-processing steps maps the resulting singular vectors back to the original channel and unit spaces to ensure that the outputs remain physically interpretable.

\paragraph{LFP pre-processing.}
Before applying \pygplaName, raw LFP data must be converted into a GPLA-ready representation. First, the LFP is filtered in the frequency band of interest. The filter bandwidth reflects a trade-off: it should be narrow enough to yield a meaningful instantaneous phase estimate, while preserving the temporal dynamics of the oscillation \cite{chavez2006proper}. The filtered signal is then transformed into a complex analytic signal using the Hilbert transform, yielding $L^{(k)}$ whose phase captures oscillatory timing and whose magnitude reflects instantaneous amplitude. These filtering and analytic-signal extraction steps are not performed by \pygplaName, because the appropriate filtering choices depend on the recording and scientific question. Instead, \pygplaName\ accepts a user-prepared complex analytic LFP signal or a real-valued phase representation in radians.

Once a complex analytic LFP signal has been supplied, \pygplaName\ can optionally whiten it to decorrelate channels. Whitening is required to justify analytical significance testing \citep{safaviUnivariateMultivariateCoupling2021} based on random matrix theory \citep[RMT, ][]{anderson2010random}; otherwise, the gPLV and spike--LFP vectors can be computed in the original channel space. When whitening is applied, GPLA outputs remain interpretable in channel space by inverting the whitening transformation.
Following \citet{safavi2023uncovering}, \pygplaName\ implements a \emph{reduced-rank whitening} procedure that avoids over-amplification of noise components \citep[i.e., small covariance eigenvalues; see ][for more details]{safavi2023uncovering} and enforces a consistent effective rank across trials. Let the concatenated analytic LFP be
\begin{equation}
L = \bigl[\, L^{(1)} \;\; L^{(2)} \;\; \cdots \;\; L^{(K)} \,\bigr] \in \mathbb{C}^{N_c \times (K T)}\,.
\end{equation}
Then, we compute an empirical covariance
\begin{equation}
\mathrm{Cov}(L) = \frac{1}{K T} L L^{H}\,,
\end{equation}
where $(\cdot)^{H}$ denotes the conjugate transpose. The effective rank $N_c^{\mathrm{eff}}$ is chosen by retaining the leading eigenmodes of $\mathrm{Cov}(L)$ that explain a fixed proportion of variance (e.g., $99\%$), and this shared value of $N_c^{\mathrm{eff}}$ is then used across trials. For each trial $k$, let
\begin{equation}
X^{(k)} \in \mathbb{C}^{N_c \times N_c^{\mathrm{eff}}}
\quad \text{and} \quad
\Lambda^{(k)} \in \mathbb{R}^{N_c^{\mathrm{eff}} \times N_c^{\mathrm{eff}}}\,
\end{equation}
denote the leading eigenvectors and eigenvalues of the single-trial covariance. The whitened LFP is then computed as
\begin{equation}
L_{w}^{(k)} = \bigl(\Lambda^{(k)}\bigr)^{-1/2} \bigl(X^{(k)}\bigr)^{H} L^{(k)}
\in \mathbb{C}^{N_c^{\mathrm{eff}} \times T}\,.
\end{equation}

Within \pygplaName, coupling analyses can be restricted to a temporal subset of samples when coupling is expected to be time-localized \citep[e.g., event-locked or transient regimes; see][]{safavi2023brain, womelsdorfModulationNeuronalInteractions2007a, Joo2018hippocampal, Lundqvist2024beta, Vinck2025predictive}. This is done by selecting an index set
\begin{equation}
\mathcal{T}_k \subseteq \{1,\dots,T\}\,,
\end{equation}
for each trial. Units can similarly be selected by retaining those whose total spike count exceeds a threshold and/or by restricting the analysis to a predefined subset
\begin{equation}
\mathcal{U} \subseteq \{1,\dots,N_u\}\,.
\end{equation}

To estimate a single coupling structure across trials within a condition, \pygplaName\ concatenates trials along time after temporal and unit selection. For equal-length selections, this yields
\begin{equation}
S \in \mathbb{R}^{|\mathcal{U}| \times (K T)},
\qquad
L \in \mathbb{C}^{N_c \times (K T)}\,,
\end{equation}
or the whitened representation
\begin{equation}
L_w \in \mathbb{C}^{N_c^{\mathrm{eff}} \times (K T)}\,,
\end{equation}
when whitening is applied. This aggregation assumes that the relevant coupling statistics are approximately stationary across the concatenated trials; when this assumption is not appropriate, GPLA can be applied separately per trial or per condition, or within sliding or event-locked windows \citep{Nouri2025}.

Lastly, when spikes and LFP are recorded from the same electrode contact, coupling estimates may be biased by shared measurement or preprocessing effects \citep{zanos2011removal}. \pygplaName\ provides a mitigation in which specified coupling entries are replaced by estimates computed from jittered surrogate spikes, reducing same-contact bias without discarding the unit or channel from the analysis \citep[see ][for more details]{safavi2023uncovering}.

\paragraph{Spike-dependent normalization.}
The choice of normalization controls how coupling estimates depend on spike count. Normalizing by $N_m^{\mathrm{tot}}$ yields a conventional PLV-type estimator when using a phase-only LFP representation. Alternatively, normalization by $\sqrt{N_m^{\mathrm{tot}}}$ produces coupling entries with approximately unit variance under the null hypothesis when the appropriate phase representation and whitening assumptions are satisfied. This unit-variance scaling underlies the analytical RMT-based significance test \citep[see ][for more details]{safaviUnivariateMultivariateCoupling2021}.

\paragraph{Post-processing.}
To make GPLA outputs interpretable in the channel and unit spaces of the recording, \pygplaName\ applies several post-processing operations after factorizing the coupling matrix:
\begin{equation}
\mathbf{C}
=
\mathbf{U}\mathbf{D}\mathbf{V}^{H},
\qquad
\mathbf{C}
\approx
d_1\mathbf{u}_1\mathbf{v}_1^{H},
\end{equation}
where $d_1$ is the gPLV and $(\mathbf{u}_1,\mathbf{v}_1)$ are the dominant LFP and spike vectors.

\textit{(i) Unwhitening of LFP vectors.}
When whitening is used, the singular LFP vectors are mapped back to the original channel space using the unwhitening operator:
\begin{equation}
\mathbf{u}^{\mathrm{unwhiten}}
=
\mathbf{W}_{\mathrm{inv}}\mathbf{u},
\qquad
\mathbf{u}^{\mathrm{normalized}}
=
\frac{\mathbf{u}^{\mathrm{unwhiten}}}
{\lVert\mathbf{u}^{\mathrm{unwhiten}}\rVert_2}.
\end{equation}
The unwhitening operator is obtained during the whitening procedure and provides a mapping from the reduced representation to the original channel coordinates.

\textit{(ii) Spike-vector rescaling.}
For the amplitude-preserving square-root normalization, spike-vector coefficients are reweighted by their corresponding spike counts and subsequently normalized:
\begin{equation}
\mathbf{v}^{\mathrm{rescaled}}
=
\mathbf{v}\oslash\sqrt{\mathbf{N}^{\mathrm{tot}}},
\qquad
\mathbf{v}^{\mathrm{normalized}}
=
\frac{\mathbf{v}^{\mathrm{rescaled}}}
{\lVert\mathbf{v}^{\mathrm{rescaled}}\rVert_2}.
\end{equation}
This compensates for the spike-count-dependent weighting introduced during construction of the coupling matrix.

\textit{(iii) Rotational phase convention.}
Because the SVD is invariant under a common complex rotation,
\begin{equation}
\mathbf{u}\mathbf{v}^{H}
=
\left(e^{-i\theta}\mathbf{u}\right)
\left(e^{-i\theta}\mathbf{v}\right)^{H},
\end{equation}
GPLA sets $\theta$ using the phase of the mean post-processed LFP vector:
\begin{equation}
\bar{\phi}_{\mathrm{LFP}}
=
\arg\left(
\frac{1}{N_c}\sum_{n=1}^{N_c}u_n
\right),
\qquad
\widetilde{\mathbf{u}}
=
\mathbf{u}e^{-i\bar{\phi}_{\mathrm{LFP}}},
\qquad
\widetilde{\mathbf{v}}
=
\mathbf{v}e^{-i\bar{\phi}_{\mathrm{LFP}}}.
\end{equation}
This centers the mean LFP-vector phase at zero and expresses the spike-vector phases relative to this convention.

Finally, \pygplaName{} can optionally normalize the gPLV by the dimensions of the coupling matrix:
\begin{equation}
\mathrm{gPLV}_{\mathrm{norm}}
=
\mathrm{gPLV}\,(N_cN_u)^{-1/2},
\end{equation}
where $N_c$ is the effective channel count after whitening, when whitening is applied \citep{safaviUnivariateMultivariateCoupling2021}.

\subsection{Package organization}

\pygplaName\ follows a layered design that mirrors the GPLA workflow from the preparation of GPLA-ready inputs to statistical assessment. The main entry point is a high-level API (\texttt{pygpla.api.gpla}) that orchestrates data preparation, coupling estimation, low-rank decomposition, and optional significance testing in one call. This keeps the default usage compact while preserving access to lower-level components for advanced analyses, modifications, and extensions.

Core numerical routines are separated by responsibility. The \texttt{pygpla.preprocessing} module handles trial concatenation, temporal and unit selection, spike-count filtering, and optional whitening of analytic LFP signals. The \texttt{pygpla.core} module implements coupling-matrix construction and singular value decomposition, including post-processing operations such as phase alignment and unwhitening-aware vector handling. Statistical procedures are collected in \texttt{pygpla.stats}, which provides surrogate-based testing \citep[e.g., multiple jittering schemes; also, see][]{grun2009data} and an analytical RMT-based test \citep{safaviUnivariateMultivariateCoupling2021}.

To support reproducibility and method validation, \pygplaName\ includes simulation utilities for generating Poisson and phase-locked spike trains, as well as paired spike--LFP data with transient coupling. The repository also provides a reproducible figure-generation script and automated tests covering individual numerical components and end-to-end workflows.

In addition, configuration dataclasses in \texttt{pygpla.config} group preprocessing, whitening, and statistical-testing parameters into structured objects, helping to document analysis settings and reduce configuration errors in complex pipelines.

\pygplaName\ accepts standard NumPy arrays without requiring a package-specific data container, facilitating integration with existing electrophysiology workflows. The core package depends only on NumPy, while SciPy is an optional dependency for documented signal-processing workflows.

\section{Demonstration of the \pygplaName\ package}

To better demonstrate the \pygplaName\ package, we apply it to two examples adapted from \citet{safavi2023uncovering}. In each case, the Python implementation handles the same
objects that arise in a real analysis: LFP samples, spike times, selected time windows,
coupling-matrix construction, singular-value decomposition, and significance testing. The
first example isolates the interpretation of gPLV and the spike vector in a single-channel
setting. The second keeps the same estimator but moves to a multichannel coupling matrix,
where the singular spectrum is used for analytical testing.

\subsection{Reading gPLV and spike-vector structure}

\begin{figure}[H]
\centering
\includegraphics[width=0.7\textwidth]{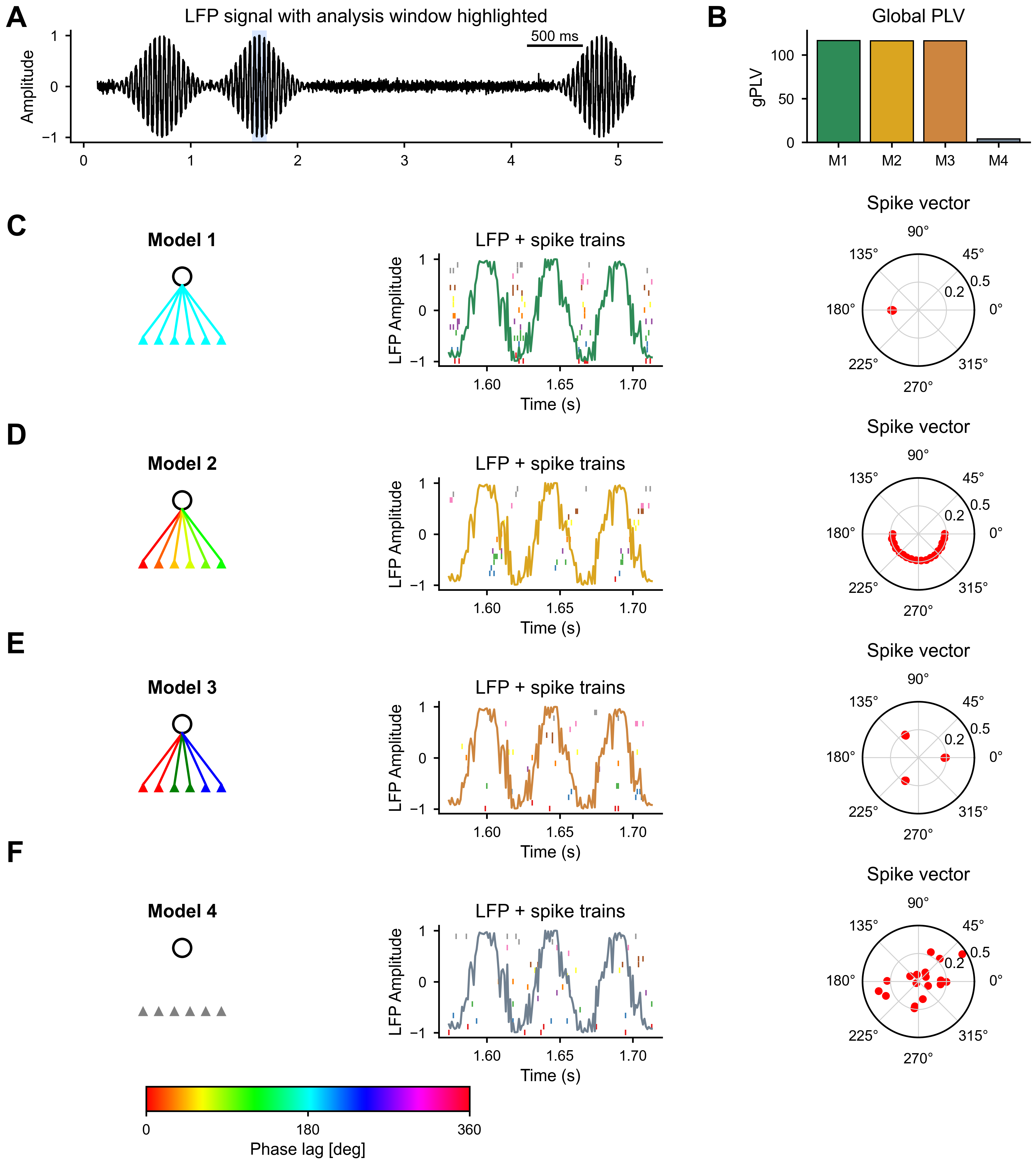}
\caption{Reproduction of the simple GPLA simulations from
\citet{safavi2023uncovering}. (A) Example transient LFP with a representative interval highlighted for visualization. (B) gPLV across four spike--LFP coupling models. (C--F) Four
coupling scenarios: globally synchronized spiking, a phase progression across
units, clustered phase locking, and uncoupled spiking. In each row, the left
panel shows a schematic of the locking structure, the middle panel shows the
LFP together with spike trains inside the selected window, and the right panel
shows the resulting spike vector in the complex plane.}
\label{fig:example-figure2}
\end{figure}

We first consider a simple setting in which a transient LFP-like oscillation modulates the
firing probability of 18 simulated spiking units. The transient signal in \mfig{fig:example-figure2}A illustrates the simulated LFP, with a representative interval highlighted for visualization. \pygplaName{} evaluates the analytic LFP at every spike time across the trials and accumulates these complex values into the spike--field coupling estimate. Because there is only one LFP channel, the coupling matrix has
 one row and one column per unit. The SVD is therefore direct to interpret: the leading singular
value gives the population-level coupling strength (gPLV), while the spike vector gives the
relative phase organization of the units within that dominant mode.

\FloatBarrier

The four coupling models in \mfig{fig:example-figure2}C--F differ only in how spikes are
locked to the oscillation. In the synchronous model, all units tend to fire in the same LFP
phase. In the sequential model, units fire with a systematic phase progression. In the clustered
model, units form a small number of phase-locked groups. The final model is an uncoupled
control in which spike times are not generated from the LFP phase. \pygplaName\ does not
receive these model identities; it only receives the spike trains and LFP. The bar plot in
\mfig{fig:example-figure2}B shows that the scalar gPLV is high for the three coupled models
and low for the uncoupled control, so it captures whether a dominant population-level
coupling exists. However, gPLV intentionally compresses the population into one scalar
value, and therefore does not distinguish synchronous, sequential, and clustered coupling by
itself.

The spike vector carries this missing organization. For the synchronous case
(\mfig{fig:example-figure2}C), the coefficients collapse to a common angle in the complex
plane, indicating that units share a preferred phase relative to the LFP. For the sequential
case (\mfig{fig:example-figure2}D), the coefficients spread along an arc, reflecting the
progressive phase lag imposed across units. For the clustered case
(\mfig{fig:example-figure2}E), the coefficients separate into a few groups, matching the
underlying subpopulations. For the uncoupled control (\mfig{fig:example-figure2}F), the
coefficients no longer form a stable phase pattern. Thus, in this example \pygplaName\ gives
two complementary readouts: the gPLV asks whether the population is coupled to the field,
and the spike vector asks how that coupling is arranged across units. This distinction is
especially important because pooling spikes across units would obscure the phase diversity
seen in the sequential and clustered models.

\subsection{Testing multichannel coupling with the singular spectrum}

\begin{figure}[h!]
\centering
\includegraphics[width=\textwidth]{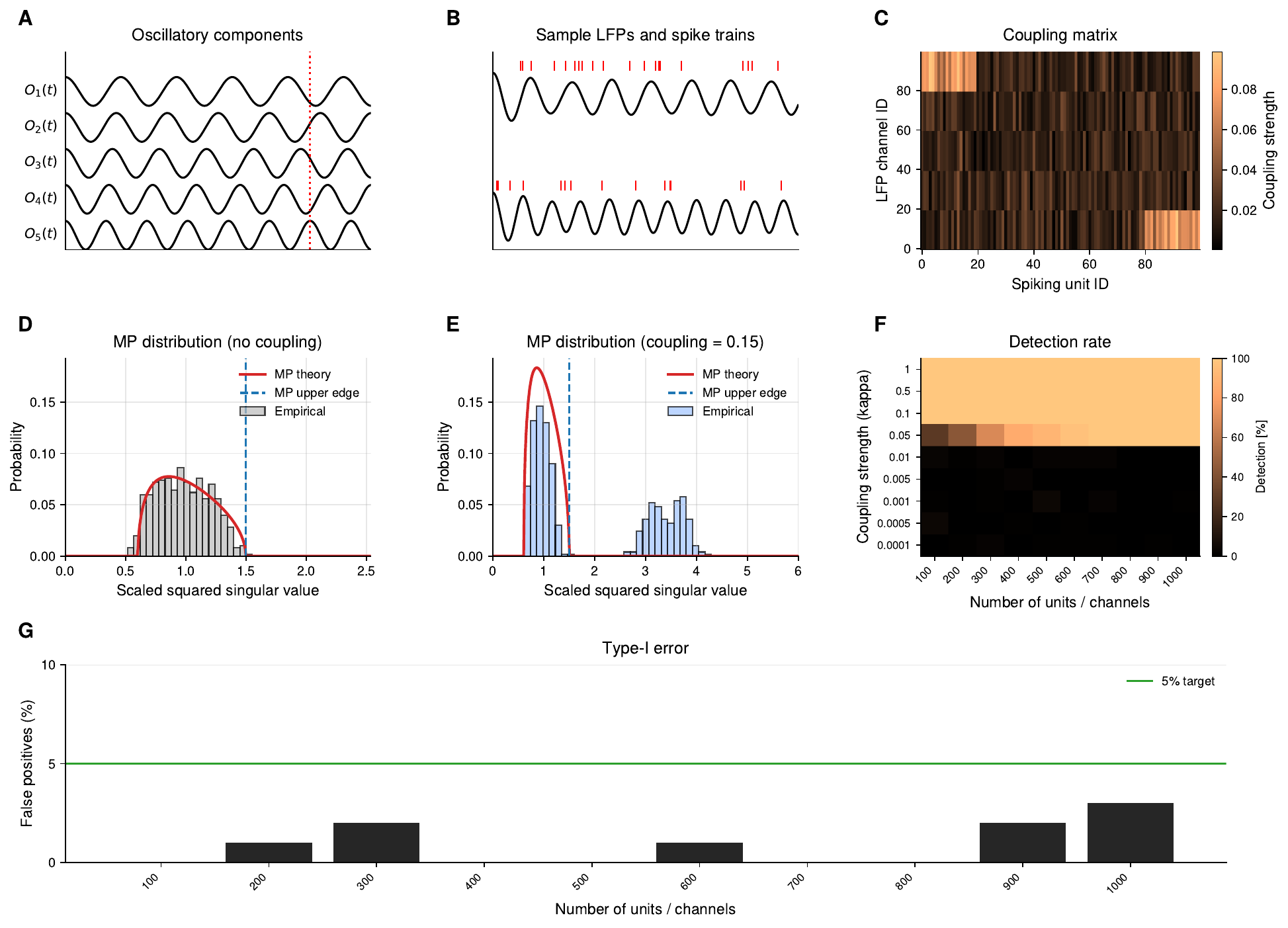}
\caption{Statistical characterization of GPLA under simulated spike--LFP coupling, computed using \pygplaName{} following the simulation paradigms of Fig.~4A--G in \citet{safavi2023uncovering}.
(A) Synthetic multichannel LFPs are generated by linearly mixing oscillatory components with distinct frequencies;
the vertical red line highlights their relative phase offsets. (B) Example LFP traces and spike trains from two coupled populations.
(C) Example spike--LFP coupling matrix from a simulation with two coupled populations.
(D--E) Empirical distributions of scaled squared singular values obtained with GPLA,
together with the theoretical Marchenko--Pastur distribution and its upper edge, for simulations (D) without coupling and (E) with coupling.
In the coupled case, signal-related singular values emerge beyond the theoretical bulk distribution.
(F) Detection rate of the RMT-based GPLA significance test as a function of coupling strength and the number of spiking units/LFP channels.
(G) Type~I error across recording dimensionalities, defined as the percentage of simulations in which significant coupling
is detected despite the absence of ground-truth coupling; the horizontal green line marks the nominal 5\% false-positive rate.}
\label{fig:example-figure4}
\end{figure}

The second example moves from a single LFP channel to a multichannel domain. Here,
several oscillatory components with different (but close) frequencies are linearly mixed to generate LFP
channels (\mfig{fig:example-figure4}A), and different unit populations are coupled to
different oscillatory components. The sample traces in \mfig{fig:example-figure4}B show the
kind of input the package receives: multichannel oscillatory activity together with spike trains
whose timing depends on specific components of that activity. For each LFP channel and each
unit, \pygplaName\ computes a complex spike-triggered coupling estimate. These entries form
a full \(N_c\times N_u\) coupling matrix rather than separate pairwise results.

The matrix in \mfig{fig:example-figure4}C illustrates why the multivariate formulation is
useful. The bright blocks indicate groups of channels and units that share an oscillatory drive:
one coupled population is associated with one part of the LFP mixture, and another
population with a different component. In the implementation, this matrix is the object passed
to the SVD. Its singular values quantify the strength of dominant coupling modes, while the
corresponding LFP and spike vectors describe which channels and units participate in those
modes and at which relative phases. The goal is therefore not to inspect every matrix entry
separately, but to extract coherent low-rank structure from the full spike--field coupling
matrix.

This example also shows how \pygplaName\ handles significance testing in high-dimensional settings.
A large singular value can arise from true coupling, but finite spike counts and many
channel--unit comparisons also produce random singular-value fluctuations. For the
analytical RMT-based test, the input should satisfy the preprocessing assumptions
required by the theory. LFP whitening can be used to reduce channel covariance, while
spike-count normalization controls the variance of coupling estimates across units.
The scaled squared singular values are then compared with the Marchenko--Pastur
distribution \citep[see ][for more details]{safaviUnivariateMultivariateCoupling2021}.
In the no-coupling simulation (\mfig{fig:example-figure4}D), the empirical
singular-value distribution remains within the theoretical bulk. When coupling is present (\mfig{fig:example-figure4}E), signal-related
singular values separate from that bulk and cross the theoretical upper edge, providing a
criterion for significant multivariate coupling.

The performance panels summarize the behavior of this test across simulation conditions.
In \mfig{fig:example-figure4}F, detection increases with coupling strength and remains strong
as the number of channels and units grows. This is the key statistical difference from running
many pairwise tests: increasing dimensionality not only increases the number of
comparisons, but it also provides a larger matrix in which coherent low-rank coupling can separate
from the null spectrum. \mfig{fig:example-figure4}G then checks the complementary failure
mode, showing the percentage of false positives when no ground-truth coupling is present.
Across the tested dimensionalities, the false-positive rate stays below the nominal 5\% target.
Together, these results show the full inferential path implemented in \pygplaName: construct
the coupling matrix, extract its dominant GPLA modes, and evaluate whether the observed
singular values are larger than expected under a dimension-aware null model.

\section{Code availability}

The source code, tests, and reproducible examples are available on
\href{https://github.com/CMC-unit/PyGPLA}{GitHub}. \pygplaName{} can be installed from
\href{https://pypi.org/project/pygpla/}{PyPI}, with documentation hosted on
\href{https://pygpla.readthedocs.io/}{Read the Docs}.

\section{Acknowledgments}

We acknowledge contributions from collaborators who provided feedback during the development of \pygplaName. S.S. acknowledges the support from the Max Planck Society and an add-on fellowship from the Joachim Herz Foundation. C.B.  acknowledges the support by the Federal Ministry of Research, Technology and Space (Bundesministerium f\"ur Forschung, Technologie und Raumfahrt; BMFTR) as part of the German Center for Child and Adolescent Health (DZKJ) under the funding code 01GL2405B.

\section{Author contribution}

Conceptualization, AK, SS;
Methodology, AK, SS;
Project administration, AK;
Software, AK;
Formal Analysis, AK, SS;
Investigation, AK;
Resources, SS;
Data Curation, AK;
Writing - Original Draft, AK, SS;
Writing - Review \& Editing: AK, CB, SS;
Visualization, AK, SS;
Supervision, SS;
Funding acquisition, CB, SS.

\section*{References}
\printbibliography[heading=none]

@article{buzsaki2012origin,
  title = {The origin of extracellular fields and currents--EEG, ECoG, LFP and spikes},
  author = {Buzsaki, Gyorgy and Anastassiou, Costas A. and Koch, Christof},
  journal = {Nature Reviews Neuroscience},
  year = {2012},
  volume = {13},
  number = {6},
  pages = {407--420},
  doi = {10.1038/nrn3241},
  pmid = {22595786}
}

@article{einevoll2013modelling,
  title = {Modelling and analysis of local field potentials for studying the function of cortical circuits},
  author = {Einevoll, Gaute T. and Kayser, Christoph and Logothetis, Nikos K. and Panzeri, Stefano},
  journal = {Nature Reviews Neuroscience},
  year = {2013},
  volume = {14},
  number = {11},
  pages = {770--785},
  doi = {10.1038/nrn3599},
  pmid = {24135696}
}

@article{hagen2016hybrid,
  title = {Hybrid Scheme for Modeling Local Field Potentials from Point-Neuron Networks},
  author = {Hagen, Espen and Dahmen, Daniel and Stavrinou, Maria L. and Linden, Henrik and Tetzlaff, Tom and van Albada, Sacha J. and Diesmann, Markus},
  journal = {Cerebral Cortex},
  year = {2016},
  volume = {26},
  number = {12},
  pages = {4461--4496},
  doi = {10.1093/cercor/bhw237},
  pmid = {27797828}
}

@article{herreras2016local,
  title = {Local Field Potentials: Myths and Misunderstandings},
  author = {Herreras, Oscar},
  journal = {Frontiers in Neural Circuits},
  year = {2016},
  volume = {10},
  pages = {101},
  doi = {10.3389/fncir.2016.00101},
  pmid = {28018180}
}

@article{zeitler2006assessing,
  title = {Assessing Neuronal Coherence with Single-Unit, Multi-Unit, and Local Field Potentials},
  author = {Zeitler, Markus and Fries, Pascal and Gielen, Stan},
  journal = {Neural Computation},
  year = {2006},
  volume = {18},
  pages = {2256--2281},
  doi = {10.1162/neco.2006.18.9.2256},
  pmid = {16846392}
}

@article{chavez2006proper,
  title = {Towards a proper estimation of phase synchronization from time series},
  author = {Chavez, M. and Besserve, M. and Adam, C. and Martinerie, J.},
  journal = {Journal of Neuroscience Methods},
  year = {2006},
  volume = {154},
  number = {1-2},
  pages = {149--160},
  doi = {10.1016/j.jneumeth.2005.12.009},
  pmid = {16445988}
}

@article{vinck2010pairwise,
  title = {The Pairwise Phase Consistency: A Bias-Free Measure of Rhythmic Neuronal Synchronization},
  author = {Vinck, Martin and van Wingerden, Marijn and Womelsdorf, Thilo and Fries, Pascal and Pennartz, Cyriel M. A.},
  journal = {NeuroImage},
  year = {2010},
  volume = {51},
  number = {1},
  pages = {112--122},
  doi = {10.1016/j.neuroimage.2010.01.073},
  pmid = {20114076}
}

@book{anderson2010random,
  title = {An Introduction to Random Matrices},
  author = {Anderson, Greg W. and Guionnet, Alice and Zeitouni, Ofer},
  publisher = {Cambridge University Press},
  year = {2010},
  series = {Cambridge Studies in Advanced Mathematics},
  volume = {118},
  pages = {xiv+492},
  isbn = {978-0-521-19452-5},
  doi = {10.1017/CBO9780511801334}
}

@article{vinck2012improved,
  title = {Improved Measures of Phase-Coupling between Spikes and the Local Field Potential},
  author = {Vinck, Martin and Battaglia, Francesco P. and Womelsdorf, Thilo and Pennartz, Cyriel M. A.},
  journal = {Journal of Computational Neuroscience},
  year = {2012},
  volume = {33},
  number = {1},
  pages = {53--75},
  doi = {10.1007/s10827-011-0374-4},
  pmid = {22187161}
}

@article{jiang2015measuring,
  title = {Measuring Directionality between Neuronal Oscillations of Different Frequencies},
  author = {Jiang, Hong and Bahramisharif, Ali and van Gerven, Marcel A. J. and Jensen, Ole},
  journal = {NeuroImage},
  year = {2015},
  volume = {118},
  pages = {359--367},
  doi = {10.1016/j.neuroimage.2015.05.044},
  pmid = {26025291}
}

@article{li2016unbiased,
  title = {Unbiased and Robust Quantification of Synchronization between Spikes and Local Field Potential},
  author = {Li, Zhixiong and Cui, Dian and Li, Xiaoli},
  journal = {Journal of Neuroscience Methods},
  year = {2016},
  volume = {269},
  pages = {33--38},
  doi = {10.1016/j.jneumeth.2016.05.004},
  pmid = {27180930}
}

@article{zarei2018introducing,
  title = {Introducing a Comprehensive Framework to Measure Spike-LFP Coupling},
  author = {Zarei, Maryam and Jahed, Maryam and Daliri, Mohammad R.},
  journal = {Frontiers in Computational Neuroscience},
  year = {2018},
  volume = {12},
  pages = {78},
  doi = {10.3389/fncom.2018.00078},
  pmid = {30374297}
}

@article{juavinett2019chronically,
  title = {Chronically Implanted Neuropixels Probes Enable High-Yield Recordings in Freely Moving Mice},
  author = {Juavinett, Ashley L. and Bekheet, Galal and Churchland, Anne K.},
  journal = {eLife},
  year = {2019},
  volume = {8},
  pages = {e47188},
  doi = {10.7554/eLife.47188},
  pmid = {31411559}
}

@article{buzsaki2004large,
  title = {Large-Scale Recording of Neuronal Ensembles},
  author = {Buzsaki, Gyorgy},
  journal = {Nature Neuroscience},
  year = {2004},
  volume = {7},
  number = {5},
  pages = {446--451},
  doi = {10.1038/nn1233},
  pmid = {15114356}
}

@article{ermentrout2001traveling,
  title = {Traveling electrical waves in cortex: insights from phase dynamics and speculation on a computational role},
  author = {Ermentrout, G. Bard and Kleinfeld, David},
  journal = {Neuron},
  year = {2001},
  volume = {29},
  number = {1},
  pages = {33--44},
  doi = {10.1016/S0896-6273(01)00178-7},
  pmid = {11182079}
}

@article{safavi2023uncovering,
  title = {Uncovering the organization of neural circuits with Generalized Phase Locking Analysis},
  author = {Safavi, Shervin and Panagiotaropoulos, Theofanis I. and Kapoor, Vishal and Ramirez-Villegas, Juan F. and Logothetis, Nikos K. and Besserve, Michel},
  journal = {PLoS Computational Biology},
  year = {2023},
  volume = {19},
  number = {4},
  pages = {e1010983},
  doi = {10.1371/journal.pcbi.1010983},
  pmid = {37011110}
}

@article{GramfortEtAl2013a,
  title = {{{MEG}} and {{EEG}} Data Analysis with {{MNE}}-{{Python}}},
  author = {Gramfort, Alexandre and Luessi, Martin and Larson, Eric and Engemann, Denis A. and Strohmeier, Daniel and Brodbeck, Christian and Goj, Roman and Jas, Mainak and Brooks, Teon and Parkkonen, Lauri and H{\"a}m{\"a}l{\"a}inen, Matti S.},
  year = {2013},
  volume = {7},
  pages = {1--13},
  doi = {10.3389/fnins.2013.00267},
  journal = {Frontiers in Neuroscience},
  number = {267}
}

@article{Nouri2025, doi = {10.21105/joss.09302}, url = {https://doi.org/10.21105/joss.09302}, year = {2025}, publisher = {The Open Journal}, volume = {10}, number = {116}, pages = {9302}, author = {Nouri, Salar and Shao, Kaidi and Safavi, Shervin}, title = {TranCIT: Transient Causal Interaction Toolbox}, journal = {Journal of Open Source Software} }

@article{friesRhythmsCognitionCommunication2015,
  title = {Rhythms for {{Cognition}}: {{Communication}} through {{Coherence}}},
  author = {Fries, P.},
  year = {2015},
  month = oct,
  volume = {88},
  pages = {220--35},
  issn = {1097-4199 (Electronic) 0896-6273 (Linking)},
  doi = {10.1016/j.neuron.2015.09.034},
  journal = {Neuron}
}

@article{womelsdorfModulationNeuronalInteractions2007a,
  title = {Modulation of Neuronal Interactions through Neuronal Synchronization},
  author = {Womelsdorf, T. and Schoffelen, J. M. and Oostenveld, R. and Singer, W. and Desimone, R. and Engel, A. K. and Fries, P.},
  year = {2007},
  month = jun,
  volume = {316},
  pages = {1609--12},
  issn = {1095-9203 (Electronic) 0036-8075 (Linking)},
  doi = {10.1126/science.1139597},
  journal = {Science},
  language = {eng}
}

@article{itoSaccaderelatedModulationsNeuronal2011,
  title = {Saccade-Related Modulations of Neuronal Excitability Support Synchrony of Visually Elicited Spikes},
  author = {Ito, Junji and Maldonado, Pedro and Singer, Wolf and Gr{\"u}n, Sonja},
  year = 2011,
  month = nov,
  journal = {Cerebral Cortex},
  volume = {21},
  number = {11},
  pages = {2482--2497},
  address = {New York, N.Y. : 1991},
  issn = {1460-2199},
  doi = {10.1093/cercor/bhr020},
  langid = {english},
  pmcid = {PMC3183421},
  pmid = {21459839}
}

@incollection{ashida2010processing,
  title = {Processing of Phase-Locked Spikes and Periodic Signals},
  author = {Ashida, Go and Wagner, Hermann and Carr, Catherine E.},
  booktitle = {Analysis of Parallel Spike Trains},
  editor = {Gr{\"u}n, Sonja and Rotter, Stefan},
  year = {2010},
  series = {Springer Series in Computational Neuroscience},
  volume = {7},
  pages = {59--74},
  publisher = {Springer},
  address = {Boston, MA},
  isbn = {978-1-4419-5675-0},
  doi = {10.1007/978-1-4419-5675-0_4}
}

@article{cruddasCorticalTravelingWaves2026,
  title = {Cortical Traveling Waves in Time and Space: {{Physics}}, Physiology, and Psychology},
  shorttitle = {Cortical Traveling Waves in Time and Space},
  author = {Cruddas, Jace and Pang, James C. and Fornito, Alex},
  date = {2026-03-18},
  journaltitle = {Neuron},
  shortjournal = {Neuron},
  volume = {114},
  number = {6},
  eprint = {41666935},
  eprinttype = {pubmed},
  pages = {985--1005},
  publisher = {Elsevier},
  issn = {0896-6273},
  doi = {10.1016/j.neuron.2025.12.019},
  url = {https://www.cell.com/neuron/abstract/S0896-6273(25)00975-4},
  urldate = {2026-04-12},
  langid = {english}
}

@article{doi:10.1137/06066638X,
author = { Troy, William C. and  Shusterman, Vladimir},
title = {Patterns and Features of Families of Traveling Waves in Large‐Scale Neuronal Networks},
journal = {SIAM Journal on Applied Dynamical Systems},
volume = {6},
number = {1},
pages = {263-292},
year = {2007},
doi = {10.1137/06066638X},

URL = { 
    
        https://doi.org/10.1137/06066638X
    
    

},
eprint = { 
    
        https://doi.org/10.1137/06066638X
    
    

}
}

@article{zhangThetaAlphaOscillations2018,
  title = {Theta and {{Alpha Oscillations Are Traveling Waves}} in the {{Human Neocortex}}},
  author = {Zhang, Honghui and Watrous, Andrew J. and Patel, Ansh and Jacobs, Joshua},
  date = {2018-06-27},
  journaltitle = {Neuron},
  shortjournal = {Neuron},
  volume = {98},
  number = {6},
  eprint = {29887341},
  eprinttype = {pubmed},
  pages = {1269-1281.e4},
  publisher = {Elsevier},
  issn = {0896-6273},
  doi = {10.1016/j.neuron.2018.05.019},
  url = {https://www.cell.com/neuron/abstract/S0896-6273(18)30417-3},
  urldate = {2026-04-12},
  langid = {english}
}

@article{mullerCorticalTravellingWaves2018,
  title = {Cortical Travelling Waves: Mechanisms and Computational Principles},
  shorttitle = {Cortical Travelling Waves},
  author = {Muller, Lyle and Chavane, Frédéric and Reynolds, John and Sejnowski, Terrence J.},
  date = {2018-05},
  journaltitle = {Nature Reviews Neuroscience},
  shortjournal = {Nat Rev Neurosci},
  volume = {19},
  number = {5},
  pages = {255--268},
  issn = {1471-003X, 1471-0048},
  doi = {10.1038/nrn.2018.20},
  url = {https://www.nature.com/articles/nrn.2018.20},
  urldate = {2026-04-12},
  langid = {english}
}

@article{dugueTravelingWavesScales2025,
  title = {Traveling Waves across Scales: {{Different}} Mechanisms but Same Canonical Computation?},
  shorttitle = {Traveling Waves across Scales},
  author = {Dugué, Laura and Chavane, Frédéric},
  editor = {Martin, Andrea E and Colgin, Laura L},
  date = {2025-11-19},
  journaltitle = {eLife},
  volume = {14},
  pages = {e106753},
  publisher = {eLife Sciences Publications, Ltd},
  issn = {2050-084X},
  doi = {10.7554/eLife.106753},
  url = {https://doi.org/10.7554/eLife.106753},
  urldate = {2026-04-12}
}

@article{dickeySingleUnitStabilityUsing2009,
  title = {Single-{{Unit Stability Using Chronically Implanted Multielectrode Arrays}}},
  author = {Dickey, Adam S. and Suminski, Aaron and Amit, Yali and Hatsopoulos, Nicholas G.},
  date = {2009-08},
  journaltitle = {Journal of Neurophysiology},
  shortjournal = {J Neurophysiol},
  volume = {102},
  number = {2},
  eprint = {19535480},
  eprinttype = {pubmed},
  pages = {1331--1339},
  issn = {0022-3077},
  doi = {10.1152/jn.90920.2008},
  url = {https://pmc.ncbi.nlm.nih.gov/articles/PMC2724357/},
  urldate = {2026-04-12},
  pmcid = {PMC2724357}
}

@article{junFullyIntegratedSilicon2017a,
  title = {Fully {{Integrated Silicon Probes}} for {{High-Density Recording}} of {{Neural Activity}}},
  author = {Jun, James J. and Steinmetz, Nicholas A. and Siegle, Joshua H. and Denman, Daniel J. and Bauza, Marius and Barbarits, Brian and Lee, Albert K. and Anastassiou, Costas A. and Andrei, Alexandru and Aydın, Çağatay and Barbic, Mladen and Blanche, Timothy J. and Bonin, Vincent and Couto, João and Dutta, Barundeb and Gratiy, Sergey L. and Gutnisky, Diego A. and Häusser, Michael and Karsh, Bill and Ledochowitsch, Peter and Lopez, Carolina Mora and Mitelut, Catalin and Musa, Silke and Okun, Michael and Pachitariu, Marius and Putzeys, Jan and Rich, P. Dylan and Rossant, Cyrille and Sun, Wei-lung and Svoboda, Karel and Carandini, Matteo and Harris, Kenneth D. and Koch, Christof and O'Keefe, John and Harris, Timothy D.},
  date = {2017-11-08},
  journaltitle = {Nature},
  shortjournal = {Nature},
  volume = {551},
  number = {7679},
  eprint = {29120427},
  eprinttype = {pubmed},
  pages = {232--236},
  issn = {0028-0836},
  doi = {10.1038/nature24636},
  url = {https://pmc.ncbi.nlm.nih.gov/articles/PMC5955206/},
  urldate = {2026-04-12},
  pmcid = {PMC5955206}
}

@article{pesaranInvestigatingLargescaleBrain2018,
  title = {Investigating Large-Scale Brain Dynamics Using Field Potential Recordings: Analysis and Interpretation},
  shorttitle = {Investigating Large-Scale Brain Dynamics Using Field Potential Recordings},
  author = {Pesaran, Bijan and Vinck, Martin and Einevoll, Gaute T. and Sirota, Anton and Fries, Pascal and Siegel, Markus and Truccolo, Wilson and Schroeder, Charles E. and Srinivasan, Ramesh},
  date = {2018-07},
  journaltitle = {Nature neuroscience},
  shortjournal = {Nat Neurosci},
  volume = {21},
  number = {7},
  eprint = {29942039},
  eprinttype = {pubmed},
  pages = {903--919},
  issn = {1097-6256},
  doi = {10.1038/s41593-018-0171-8},
  url = {https://pmc.ncbi.nlm.nih.gov/articles/PMC7386068/},
  urldate = {2026-04-12},
  pmcid = {PMC7386068}
}

@article{behrad2025fast,
  title = {Fast dynamical similarity analysis},
  author = {Behrad, Arman and Ostrow, Mitchell and Fakharian, Mohammad Taha and Fiete, Ila and Beste, Christian and Safavi, Shervin},
  journal = {arXiv},
  year = {2025},
  volume = {2511.22828},
  pages = {1--31},
  doi = {10.48550/arXiv.2511.22828}
}

@article{freeman2015open,
  title = {Open source tools for large-scale neuroscience},
  author = {Freeman, Jeremy},
  journal = {Current Opinion in Neurobiology},
  year = {2015},
  volume = {32},
  pages = {156--163},
  doi = {10.1016/j.conb.2015.04.002}
}

@article{akam2022open,
  title = {Open-source, Python-based, hardware and software for controlling behavioural neuroscience experiments},
  author = {Akam, Thomas and Lustig, Andy and Rowland, James M. and Kapanaiah, Sampath K. T. and Esteve-Agraz, Joan and Panniello, Mariangela and M{\'a}rquez, Cristina and Kohl, Michael M. and K{\"a}tzel, Dennis and Costa, Rui M. and Walton, Mark E.},
  journal = {eLife},
  year = {2022},
  volume = {11},
  pages = {e67846},
  doi = {10.7554/eLife.67846}
}

@article{goodman2008brian,
  title = {Brian: A simulator for spiking neural networks in Python},
  author = {Goodman, Dan F. M. and Brette, Romain},
  journal = {Frontiers in Neuroinformatics},
  year = {2008},
  volume = {2},
  pages = {5},
  doi = {10.3389/neuro.11.005.2008}
}

@article{ince2009python,
  title = {Python for information theoretic analysis of neural data},
  author = {Ince, Robin A. A. and Petersen, Rasmus S. and Swan, Daniel C. and Panzeri, Stefano},
  journal = {Frontiers in Neuroinformatics},
  year = {2009},
  volume = {3},
  pages = {4},
  doi = {10.3389/neuro.11.004.2009}
}

@article{krause2014expyriment,
  title = {Expyriment: A Python library for cognitive and neuroscientific experiments},
  author = {Krause, Florian and Lindemann, Oliver},
  journal = {Behavior Research Methods},
  year = {2014},
  volume = {46},
  pages = {416--428},
  doi = {10.3758/s13428-013-0390-6}
}

@article{makowski2021neurokit2,
  title = {{NeuroKit}2: A Python toolbox for neurophysiological signal processing},
  author = {Makowski, Dominique and Pham, Tam and Lau, Zen J. and Brammer, Jan C. and Lespinasse, Fran{\c{c}}ois and Pham, Hung and Sch{\"o}lzel, Christopher and Chen, S. H. Annabel},
  journal = {Behavior Research Methods},
  year = {2021},
  volume = {53},
  pages = {1689--1696},
  doi = {10.3758/s13428-020-01516-y}
}

@article{muller2015python,
  title = {Python in neuroscience},
  author = {Muller, Eilif and Bednar, James A. and Diesmann, Markus and Gewaltig, Marc-Oliver and Hines, Michael and Davison, Andrew P.},
  journal = {Frontiers in Neuroinformatics},
  year = {2015},
  volume = {9},
  pages = {11},
  doi = {10.3389/fninf.2015.00011}
}

@article{peirce2009generating,
  title = {Generating stimuli for neuroscience using {PsychoPy}},
  author = {Peirce, Jonathan W.},
  journal = {Frontiers in Neuroinformatics},
  year = {2009},
  volume = {2},
  pages = {10},
  doi = {10.3389/neuro.11.010.2008}
}

@article{tennoe2018uncertainpy,
  title = {Uncertainpy: A Python toolbox for uncertainty quantification and sensitivity analysis in computational neuroscience},
  author = {Tenn{\o}e, Simen and Halnes, Geir and Einevoll, Gaute T.},
  journal = {Frontiers in Neuroinformatics},
  year = {2018},
  volume = {12},
  pages = {49},
  doi = {10.3389/fninf.2018.00049}
}

@article{viejo2023pynapple,
  title = {Pynapple, a toolbox for data analysis in neuroscience},
  author = {Viejo, Guillaume and Levenstein, Daniel and Skromne Carrasco, Sofia and Mehrotra, Dhruv and Mahallati, Sara and Vite, Gilberto R. and Denny, Henry and Sjulson, Lucas and Battaglia, Francesco P. and Peyrache, Adrien},
  journal = {eLife},
  year = {2023},
  volume = {12},
  pages = {RP85786},
  doi = {10.7554/eLife.85786.3}
}

@article{zeraati2022flexible,
  title = {A flexible Bayesian framework for unbiased estimation of timescales},
  author = {Zeraati, Roxana and Engel, Tatiana A. and Levina, Anna},
  journal = {Nature Computational Science},
  year = {2022},
  volume = {2},
  pages = {193--204},
  doi = {10.1038/s43588-022-00214-3}
}

@article{Vinck2025predictive,
  title = {Large-scale interactions in predictive processing: Oscillatory versus transient dynamics},
  author = {Vinck, Martin and Uran, Cem and Dowdall, Jarrod R. and Rummell, Brian and Canales-Johnson, Andres},
  journal = {Trends in Cognitive Sciences},
  year = {2025},
  volume = {29},
  number = {2},
  pages = {133--148},
  doi = {10.1016/j.tics.2024.09.013}
}

@article{zanos2011removal,
  title = {Removal of spurious correlations between spikes and local field potentials},
  author = {Zanos, Theodoros P. and Mineault, Patrick J. and Pack, Christopher C.},
  journal = {Journal of Neurophysiology},
  year = {2011},
  volume = {105},
  number = {1},
  pages = {474--486},
  doi = {10.1152/jn.00642.2010}
}

@article{grun2009data,
  title = {Data-driven significance estimation for precise spike correlation},
  author = {Gr{\"u}n, Sonja},
  journal = {Journal of Neurophysiology},
  year = {2009},
  volume = {101},
  number = {3},
  pages = {1126--1140},
  doi = {10.1152/jn.00093.2008}
}

@article{safavi2023brain,
  title = {Brain as a complex system, harnessing systems neuroscience tools \& notions for an empirical approach},
  author = {Safavi, Shervin},
  journal = {arXiv},
  year = {2023},
  volume = {2312.13478},
  doi = {10.48550/arXiv.2312.13478}
}

@article{safaviUnivariateMultivariateCoupling2021,
  title = {From univariate to multivariate coupling between continuous signals and point processes: A mathematical framework},
  author = {Safavi, Shervin and Logothetis, Nikos K. and Besserve, Michel},
  journal = {Neural Computation},
  year = {2021},
  volume = {33},
  number = {7},
  pages = {1751--1817},
  doi = {10.1162/neco_a_01389}
}

@article{dwarakanath2023bistability,
  title = {Bistability of prefrontal states gates access to consciousness},
  author = {Dwarakanath, Abhilash and Kapoor, Vishal and Werner, Joachim and Safavi, Shervin and Fedorov, Leonid A. and Logothetis, Nikos K. and Panagiotaropoulos, Theofanis I.},
  journal = {Neuron},
  year = {2023},
  volume = {111},
  number = {10},
  pages = {1666--1683.e4},
  doi = {10.1016/j.neuron.2023.02.027}
}

@article{Fries2005,
  title = {A mechanism for cognitive dynamics: Neuronal communication through neuronal coherence},
  author = {Fries, Pascal},
  journal = {Trends in Cognitive Sciences},
  year = {2005},
  volume = {9},
  number = {10},
  pages = {474--480},
  doi = {10.1016/j.tics.2005.08.011}
}

@article{Joo2018hippocampal,
  author  = {Hannah R. Joo and Loren M. Frank},
  title   = {The hippocampal sharp wave–ripple in memory retrieval for immediate use and consolidation},
  journal = {Nature Reviews Neuroscience},
  year    = {2018},
  volume  = {19},
  number  = {12},
  pages   = {744--757},
  month   = dec,
  doi     = {10.1038/s41583-018-0077-1},
  url     = {https://www.nature.com/articles/s41583-018-0077-1}
}

@article{Lundqvist2024beta,
  author  = {Lundqvist, Mikael and Miller, Earl K. and Nordmark, Jonatan and Liljefors, Johan and Herman, Pawel},
  title   = {Beta: Bursts of cognition},
  journal = {Trends in Cognitive Sciences},
  year    = {2024},
  month   = apr,
  doi     = {10.1016/j.tics.2024.03.010},
  url     = {https://www.sciencedirect.com/science/article/pii/S1364661324000779}
}

@Article{stevenson2011HowAdvancesNeural,
  author    = {Stevenson, Ian H. and Kording, Konrad P.},
  title     = {How Advances in Neural Recording Affect Data Analysis},
  doi       = {10.1038/nn.2731},
  issn      = {1546-1726},
  language  = {en},
  number    = {2},
  pages     = {139-142},
  volume    = {14},
  copyright = {2011 Nature Publishing Group},
  journal   = {Nature Neuroscience},
  month     = feb,
  year      = {2011},
}

@article{Newman2025ONIX,
  author  = {Newman, Jonathan P. and Zhang, Jie and Cuevas-L{\'o}pez, Aar{\'o}n
             and Miller, Nicholas J. and Honda, Takato
             and van der Goes, Marie-Sophie H.
             and Leighton, Alexandra H. and Carvalho, Filipe
             and Lopes, Gon{\c{c}}alo and Lakunina, Anna
             and Siegle, Joshua H. and Harnett, Mark T.
             and Wilson, Matthew A. and Voigts, Jakob},
  title   = {ONIX: a unified open-source platform for multimodal neural recording and perturbation during naturalistic behavior},
  journal = {Nature Methods},
  year    = {2025},
  volume  = {22},
  pages   = {187--192},
  doi     = {10.1038/s41592-024-02521-1},
  url     = {https://doi.org/10.1038/s41592-024-02521-1}
}

@article{Stringer2024AnalysisMethods,
  author    = {Carsen Stringer and Marius Pachitariu},
  title     = {Analysis methods for large-scale neuronal recordings},
  journal   = {Science},
  year      = {2024},
  volume     = {386},
  number     = {6722},
  month      = nov,
  doi        = {10.1126/science.adp7429},
  url        = {https://www.science.org/doi/10.1126/science.adp7429}
}

@article{Siegle2026LargeScaleElectrophysiology,
  author  = {Siegle, Joshua H. and Steinmetz, Nicholas A.},
  title   = {Large-scale electrophysiology at single-spike resolution},
  journal = {Nature Reviews Neuroscience},
  year    = {2026},
  volume  = {27},
  pages   = {395--415},
  doi     = {10.1038/s41583-026-01042-4},
  url     = {https://doi.org/10.1038/s41583-026-01042-4}
}

\end{document}